\documentclass{article}

\usepackage{epsfig} 
\usepackage{amsmath}
\usepackage{amsfonts}
\usepackage{graphicx}
\usepackage{cite}

\usepackage{amssymb}

\usepackage{epsf,epsfig}

 \usepackage{graphicx}
\usepackage{bm}
\usepackage{amsfonts}
\usepackage{amsmath}

\usepackage{graphicx}
\usepackage{epsfig}
\usepackage{amsfonts}
\usepackage{a4}
\usepackage{amsmath}

\newcommand{\vphi}{\varphi}

\newcommand{\ee}{\end{equation}}
\newcommand{\eea}{\end{eqnarray}}
\newcommand{\be}{\begin{equation}}
\newcommand{\bea}{\begin{eqnarray}}

\def\dalemb#1#2{{\vbox{\hrule height .#2pt
        \hbox{\vrule width.#2pt height#1pt \kern#1pt
                \vrule width.#2pt}
        \hrule height.#2pt}}}

\def\0{{\sst{(0)}}}
\def\1{{\sst{(1)}}}
\def\2{{\sst{(2)}}}
\def\3{{\sst{(3)}}}
\def\4{{\sst{(4)}}}
\def\5{{\sst{(5)}}}
\def\6{{\sst{(6)}}}
\def\7{{\sst{(7)}}}
\def\8{{\sst{(8)}}}

    \let\p=\pi

 \def\bd{\begin{document}} \def\ed{\end{document}}
\def\ds{\documentstyle} \let\fr=\frac \let\bl=\bigl \let\br=\bigr
\let\Br=\Bigr \let\Bl=\Bigl
\let\bm=\bibitem
\let\na=\nabla
\let\pa=\partial \let\ov=\overline
\def\fft#1#2{{#1 \over #2}}

    \numberwithin{equation}{section}

\begin{document}

\title{\bf \Large 
Static black holes with axial  symmetry  
\\
in  asymptotically AdS$_4$ spacetime}

\author{{\large Olga Kichakova}$^{1,2}$, {\large  Jutta Kunz}$^{1}$,
{\large Eugen Radu}$^{3}$
and {\large Yasha Shnir}$^{1,4,5}$
\\
\\
 {\small  $^{1}$Institut f\"ur Physik, Universit\"at Oldenburg, Postfach 2503
D-26111 Oldenburg, Germany}
\\
{\small $^{2}$Department of Mathematics and Statistics, University of Massachusetts,
Amherst, Massachusetts 01003-4515, USA}
\\
 {\small  $^{3}$Departamento de Fisica da Universidade de Aveiro and CIDMA,}
\\
 {\small 
 Campus de Santiago, 3810-183 Aveiro, Portugal}
\\
{\small $^{4}$Department of Theoretical Physics and Astrophysics, BSU, Minsk, Belarus}
\\
{\small $^{5}$BLTP, JINR, Dubna, Russia}
 }
\date{\today}

\maketitle

 \begin{abstract} 
 The known static electro-vacuum black holes in a globally AdS$_4$ background
 have an event horizon which is geometrically a round sphere.
In this work we argue that the situation is different in  models with matter fields possessing 
 an explicit dependence on the azimuthal angle $\varphi$,
 which, however, does not manifest at the level of the energy-momentum tensor.
 As a result, the full solutions are axially symmetric only, possessing a single (timelike) Killing vector field.
 Explicit examples of such 
 static black holes are constructed
 in Einstein--(complex) scalar field and Einstein--Yang-Mills theories.
The basic properties of these solutions are discussed, 
looking for generic features.
For example, we notice that the horizon  has an oblate spheroidal shape for solutions with a scalar field
and a prolate one for black holes with Yang-Mills fields.
The deviation from sphericity of the horizon geometry
manifests itself in the holographic stress-tensor.
Finally, based on the results
obtained in the probe limit,
we conjecture the existence in Einstein-Maxwell theory  of static black holes with axial symmetry only.
  \end{abstract}


\section{Introduction}
The asymptotically flat black holes (BHs) in $d=4$ spacetime dimensions
are rather special objects.
In the electro-vacuum case, the spectrum of physically interesting solutions
is strongly constrained by the uniqueness theorems \cite{Chrusciel:2012jk}.
Moreover,  the topology of
a spatial section of the  event horizon of a stationary BH is necessarily that of a
two-sphere, $S^2$ 
\cite{HE,Friedman:1993ty}.
However, 
the situation changes when allowing for a negative cosmological constant $\Lambda<0$,
in which case the natural background of a gravity theory corresponds to anti-de Sitter (AdS) spacetime.
First, to our best knowledge, 
the uniqueness of the BHs in the electro-vacuum theory has not been rigorously established.
Moreover, for AdS BHs, the topology of a spatial section of the  event horizon
is no longer restricted to be $S^2$, 
solutions  replacing the  
two-sphere by a two-dimensional space  of negative or vanishing curvature
being considered by many authors, see $e.g.$ Ref. \cite{Mann:1997iz}. 
These {\it `topological BHs'} approach asymptotically a 
$locally$ AdS background, being seminal to recent developments in BH physics,
in particular in the context of the AdS/CFT conjecture \cite{Maldacena:1997re,Witten:1998qj}.

However, the picture can be more complicated also for solutions with a spherical
horizon topology, which approach at infinity a globally AdS background.
As argued in this work, when allowing for a sufficiently general matter content, one finds 
BHs which are static and {\it axially symmetric} only.
 This is achieved by
allowing for 
matter fields that  do not share the symmetries with the spacetime they live in \cite{Smolic:2015txa}.
More specifically,  the matter fields 
depend on the azimuthal angle $\varphi$, a dependence which, however, does not manifest at the level of the energy-momentum tensor.
As a result, the configurations are no longer spherically symmetric,
and the full solutions (geometry $+$ matter fields) 
possess a single Killing vector  $\partial/\partial t$ (with $t$ the time coordinate).
For $\Lambda=0$ ($i.e.$ for configurations in a Minkowski spacetime background),
the existence of BHs with a single Killing vector field 
was studied some time ago for matter fields of Yang-Mills type
\cite{Kleihaus:1997ic},
whereas more recently models with scalar fields only
attracted attention,
as discussed $e.g.$ in
\cite{Kleihaus:2013tba}.
In particular, there are also spinning BHs,
which possess a single Killing vector,
like the spinning BHs with Yang-Mills hair
\cite{Kleihaus:2000kg}.
In the case of scalar fields,
as first realized in \cite{Dias:2011at}
(based on an Ansatz proposed in \cite{Hartmann:2010pm}),
these BHs may possess a single $helical$ Killing vector
\cite{Herdeiro:2014goa,Kleihaus:2015iea}.

The main purpose of this work is to investigate
static BHs with axial symmetry in asymptotically AdS$_4$ spacetime.
As expected, such solutions retain many of the properties 
of their $\Lambda=0$ counterparts.
In particular, they possess a topologically $S^2$ horizon, 
which, however, is not a round sphere.
But the asymptotically AdS$_4$ BHs possess new features as well.

The paper starts by proposing in Section 2 a general framework to describe the general properties of such configurations.
In Section 3 we employ this formalism and
consider two explicit examples of static axially symmetric BHs, namely in 
 Einstein--(complex) scalar field and Einstein--Yang-Mills (EYM) theories.
There we do not aim at a systematic study of these solutions, 
 looking instead for generic properties induced by a static non-spherical horizon.
 For example, rather unexpectedly, the horizon deformation of the solutions is different in these two cases,
the event horizon having an oblate spheroid shape for Einstein--scalar field BHs
and a prolate one for EYM BHs.
Moreover, the deviation from sphericity is manifesting also in the holographic stress-tensor. 

The general results are compiled in Section 4,
together with possible avenues for future research.
There we also speculate about the possible existence 
of  static BHs with axial symmetry only in the Einstein-Maxwell theory.
This conjecture is based on the  results 
 found in the Appendix for Maxwell fields in a fixed 
Schwarzschild-AdS (SAdS) BH background. 
The Appendix proposes a  discussion of the solutions of different types of
matter field equations in the  probe limit.
There, apart from the Maxwell case, we show the existence of static 
solutions for 
(non-linear--)Klein-Gordon and Yang-Mills equations
in a SAdS background. 
In both cases, the basic properties of
 the matter fields in the presence of backreaction are already present.

\section{A general framework}

\subsection{The problem and basic setup}
We  consider a general model describing Einstein gravity with a cosmological constant
$\Lambda=-3/L^2$ coupled with a set of matter fields $\Psi$ with a Lagrangean density
${\cal L}_{m}(\Psi)$:
\begin{eqnarray}
\label{general-model}
S=   \int_{\cal M} d^4 x\sqrt{-g}
 \left(
 \frac{1}{16\pi G}(R+\frac{6}{L^2})+{\cal L}_{m}(\Psi)
 \right) .
 \end{eqnarray}
The Einstein equations are found from the variation of (\ref{general-model}) 
with respect to the metric,
 \begin{eqnarray}
\label{Einstein-eqs}
R_{\mu\nu} =-\Lambda g_{\mu \nu}+ 8\pi G (T_{\mu\nu}-\frac{1}{2}Tg_{\mu \nu})~,
\end{eqnarray}
 with the energy-momentum tensor of the matter fields
\begin{eqnarray}
\label{Tmunu}
T_{\mu\nu}=-\frac{2}{\sqrt{-g}}\frac{\delta (\sqrt{-g} {\cal L}_{m})}{\delta g^{\mu\nu}}~.
\end{eqnarray}
Apart from (\ref{Einstein-eqs}), one should also consider the matter field equations, which  
are found from the variation of the action with respect to $\Psi$.

The solutions in this work approach asymptotically an
 AdS$_4$ background, which is written in global coordinates as
\begin{eqnarray}
\label{AdS}
ds^2=\frac{dr^2}{N(r)}+r^2(d\theta^2+\sin^2\theta d\varphi^2)-
N(r)dt^2,~~{\rm with}~~N(r)=1+\frac{r^2}{L^2}.
 \end{eqnarray}
In the above relations, $(r,t)$ are the radial and time coordinates, respectively  (with $0\leq r<\infty$ and $-\infty<t<\infty$), while
$\theta$ and $\varphi$ are angular coordinates  with the usual range, parametrizing the two dimensional sphere $S^2$.

Our choice in this work in solving the eqs.~(\ref{Einstein-eqs}) was to employ the Einstein-De Turck (EDT) approach.
This approach has been proposed in  
\cite{Headrick:2009pv,Adam:2011dn,Wiseman:2011by},
and has become recently a standard tool in the treatment of 
numerical problems in general relativity
 which result in partial differential equations.   
This scheme has the advantage of not fixing $apriori$ a metric gauge,
yielding elliptic partial differential equations
 (for a review, see the recent reference
\cite{Dias:2015nua}).

In the EDT approach, 
instead of the Einstein eqs.~(\ref{Einstein-eqs}),
 one solves the so-called  EDT  equations
\begin{eqnarray}
\label{EDT}
R_{\mu\nu}-\nabla_{(\mu}\xi_{\nu)}=-\Lambda g_{\mu \nu}+ 8\pi G (T_{\mu\nu}-\frac{1}{2}Tg_{\mu \nu})~,~~
{\rm with}~~\xi^\mu=g^{\nu\rho}(\Gamma_{\nu\rho}^\mu-\bar \Gamma_{\nu\rho}^\mu),
\end{eqnarray} 
where $\Gamma_{\nu\rho}^\mu$ is the Levi-Civita connection associated to $g_{\mu\nu}$;
also, a  reference metric $\bar g$ is introduced (with the
same boundary conditions as the metric $g$),
$\bar \Gamma_{\nu\rho}^\mu$ being the corresponding Levi-Civita connection \cite{Wiseman:2011by}.
Solutions to (\ref{EDT}) solve the Einstein equations
iff $\xi^\mu \equiv 0$ everywhere on
${\cal M}$, a condition which is verified from  the numerical output.

The configurations we are interested are static and axially symmetric,
being constructed within the 
following metric Ansatz:
\begin{eqnarray}
\label{metric}
ds^2=f_1(r,\theta)\frac{dr^2}{N(r)}+S_1(r,\theta)(g(r) d\theta+S_2(r,\theta)dr)^2
+f_2(r,\theta)g(r)^2\sin^2\theta d\varphi^2-\frac{r^2}{g(r)^2}f_0(r,\theta)N(r)dt^2,
\end{eqnarray}
with five unknown functions $f_0,~f_1,~f_2,~S_1,~S_2$
and two background functions $N(r),g(r)$ which are fixed by the choice of
the reference metric $\bar g$.

A sufficiently general choice for the 
 reference metric which is used in this paper is the one corresponding to a Reissner-Nordstr\"om-AdS (RNAdS) spacetime with\footnote{The usual
 parametrization of the RNAdS metric is recovered by taking $r\to \sqrt{r^2+r_H^2}$.}
\begin{eqnarray}
\nonumber
&&
S_1=f_1=f_2=f_0=1,~S_2=0,~~{\and}
\\
\nonumber
&&
g(r)=\sqrt{r^2+r_H^2},~~
N(r)=\frac{g(r)}{g(r)+r_H}
\left (
1+\frac{1}{L^2}(r^2+r_H(2r_H+g(r))
-\frac{Q^2}{g(r) r_H}
\right).
\end{eqnarray}
Apart from the AdS length scale $L$, this reference metric contains two other input constants,
$r_H \geq 0$ and $Q^2\geq 0$.

\subsection{The boundary conditions  for the metric functions}
The behaviour of the metric potentials $g_{\mu\nu}$
on the boundaries of the domain of integration is universal, being recovered
for any matter content\footnote{For both scalar and Yang-Mills cases, 
we have verified the existence of approximate solutions compatible with the boundary conditions
(\ref{bc-eh}),
(\ref{bc-inf}),
(\ref{bc-axis}) together with the corresponding ones for the matter fields
($e.g.$
for the near horizon case, $r\to 0$, one takes a power series in $r$ etc.).
However, the corresponding expressions are complicated and we have decided 
to not include them here.}.
All solutions in this work possess a nonextremal horizon located at $r=0$,
the following boundary conditions being imposed there
 \begin{eqnarray}
\label{bc-eh}
\partial_r f_1\big|_{r=0}=\partial_r f_2\big|_{r=0}=\partial_r f_0\big|_{r=0}
=\partial_r S_1\big|_{r=0}= S_2\big|_{r=0}=0.
\end{eqnarray} 
There are also the supplementary conditions
\begin{eqnarray}
\nonumber
 f_1\big|_{r=0}=  f_0\big|_{r=0},~~
 \partial_r S_2\big|_{r=0}= 0,
\end{eqnarray} 
which are used to verify the accuracy of solutions.
At infinity we impose
\begin{eqnarray}
\label{bc-inf}
 f_1\big|_{r=\infty}=  f_2\big|_{{r=\infty}}=  f_0\big|_{r=\infty}=S_1\big|_{r=\infty}=1,~S_2\big|_{r=\infty}=0,
\end{eqnarray}
such that the global AdS$_4$ background (\ref{AdS}) is approached.
More precisely, the matter fields considered in this work decay fast enough at infinity,
such that the far field behaviour of the unknown functions which enter the line element (\ref{metric}) is
\begin{eqnarray}
\nonumber
&&
f_0=1+\frac{f_{03}(\theta)}{r^3}+ O(1/r^4),~~f_1= O(1/r^4),~~
 f_2=1+\frac{f_{23}(\theta)}{r^3}+ O(1/r^4),
 \\
 \label{inf}
&&
S_1=1+\frac{s_{13}(\theta)}{r^3}+ O(1/r^4),~~
 S_2= O(1/r^5),
\end{eqnarray}
with $f_{03}(\theta)$, $f_{23}(\theta)$, $s_{13}(\theta)$
functions fixed by the numerics. These functions satisfy the relations
\begin{eqnarray}
\nonumber
f_{03}+f_{23}+s_{13}=0,~~\cos\theta (s_{13}-f_{23})=\sin\theta \frac{d}{d\theta}(f_{03}+f_{23}-s_{13}).
\end{eqnarray}

The boundary conditions on the symmetry axis, $\theta=0,\pi$ are
\begin{eqnarray}
\label{bc-axis}
\partial_\theta f_1\big|_{\theta=0,\pi}=\partial_\theta f_2\big|_{\theta=0,\pi}=
\partial_\theta f_0\big|_{\theta=0,\pi}=\partial_\theta S_1\big|_{\theta=0,\pi}=S_2\big|_{\theta=0,\pi}=0.
\end{eqnarray}
Moreover, all configurations in this work are
symmetric $w.r.t.$ a reflection in the 
equatorial plane, which implies
\begin{eqnarray}
\nonumber
\partial_\theta f_1\big|_{\theta=\pi/2}=\partial_\theta f_2\big|_{\theta=\pi/2}=
\partial_\theta f_0\big|_{\theta=\pi/2}=\partial_\theta S_1\big|_{\theta=\pi/2}=S_2\big|_{\theta=\pi/2}=0, 
\end{eqnarray}
such that we need to consider the solutions only in the region $0\leq \theta \leq \pi/2$.

\subsection{Quantities of interest}
\subsubsection{Horizon properties}

Starting with the horizon quantities, we note that the
solutions have an event horizon of spherical topology, which, for our 
formulation of the problem is located at $r=0$.
The induced metric on  a spatial section of the  event horizon is
\begin{eqnarray}
\label{metric-horizon}
d\sigma^2=r_H^2\big ( S_1(0,\theta)d\theta^2+f_2(0,\theta)\sin^2\theta d\varphi^2 \big),
 \end{eqnarray} 
 with $S_1(0,\theta)$, $f_2(0,\theta)$ strictly positive functions.
 Geometrically, however, the horizon is a squashed sphere.
 This can be seen by evaluating the circumference of the horizon along
the equator, 
\begin{equation}
L_e=2 \pi r_H \sqrt{f_2(0,\pi/2)} \ ,
\end{equation}
 and comparing it with the circumference of the horizon along the poles, 
\begin{equation}
L_p=2 r_H \int_0^\pi d\theta \sqrt{S_1(0,\theta)}  \ .
\end{equation} 
Then we can define an `excentricity' of the solutions,
\begin{eqnarray}
\label{exc}
\epsilon=\frac{L_e}{L_p} ,
\end{eqnarray}
with
$\epsilon>1$
in the oblate case, $\epsilon<1$
for a prolate horizon and
 $\epsilon=1$ in the spherical limit.
 
Futher insight on the horizon properties is found by 
considering its isometric embedding in a three dimensional flat space\footnote{
Note that in principle not all $d=2$ surfaces can be embedded isometrically in a $d=3$
Euclidean space. 
However, this was the case for all solutions in this work which were investigated from this direction.},
with $d\sigma_3^2=dx^2+dy^2+dz^2$.
This is achieved by taking 
$x=F(\theta) \cos\varphi$, 
$y=F(\theta) \sin\varphi$,
$z=G(\theta)$,
where
$F(\theta)=r_H\sin \theta \sqrt{f_2 }|_{r=0}$
and 
$G'(\theta)=r_H\sqrt{S_1 - \frac{1}{4f_2}(2\cos\theta f_2 +\sin \theta \frac{\partial f_2}{\partial \theta})^2}\big|_{r=0}$.
Based on that, one can define an equator radius $R_e=f(\pi/2)$ and a polar one $R_p=g(0)$.
We have found that the ratio $R_e/R_p$
has the same behaviour as $L_e/L_p$;
in particular, it takes close (but not equal) values to it.

The horizon area of a BH is given by
\begin{eqnarray}
\label{AH}
A_H= 2\p r_H^2 \int_0^\pi 
d\theta \sin\theta
\sqrt{f_2(0,\theta)S_1(0,\theta)},
\end{eqnarray} 
with the entropy $S=A_H/(4 G)$.
Finally, the Hawking temperature
$T_H={\kappa}/({2\pi})$ (where $\kappa$ is the surface gravity)
  is given by\footnote{From (\ref{temp}), one can see that in the EDT approach,
the Hawking  temperature does not follow from the numerical output, being an input parameter.}
\begin{eqnarray}
\label{temp}
T_H=\frac{1}{4\pi r_H}
\left(1-\frac{Q^2}{r_H^2}+\frac{3r_H^2}{L^2}
\right).
\end{eqnarray} 
%
 
\subsubsection{The mass and the  holographic stress tensor}
There are also a number of quantities 
defined in the far field.
The mass of the solutions  with the asymptotics (\ref{inf}),
as computed according to both the quasilocal  prescription in 
\cite{Balasubramanian:1999re} 
and the Ashtekar-Das one
\cite{Ashtekar:1999jx}
is given by
\begin{eqnarray}
\nonumber
M=\frac{1}{8 G } 
\int_0^\pi d\theta 
\left(
\frac{3}{L^2}(f_{23}(\theta)+s_{13}(\theta))
+2(\frac{Q^2}{r_H}+r_H+\frac{r_H^3}{L^2})
\right)
\sin \theta,
\end{eqnarray}
with $f_{23}(\theta)$, $s_{13}(\theta)$
the functions which enter the large-$r$ asympotics (\ref{inf}).

It is also of interest to evaluate the holographic stress tensor.
In order to extract it,
we first transform the (asymptotic metric) into Fefferman--Graham coordinates, by using a new radial coordinate $z$, with
\begin{eqnarray}
\label{cd1}
 r=\frac{L^2}{z}-\frac{2r_H^2+L^2}{4L^2}z+\frac{r_H^4+(Q^2+r_H^2)L^2}{6r_HL^4}z^2.
\end{eqnarray}
 In these coordinates, the line element can be expanded around $z=0$  ($i.e.$ as $r\to \infty$) 
in the standard form
\begin{eqnarray}
\label{FG1}
ds^2=\frac{L^2}{z^2}
\bigg [
dz^2+\big(g_{(0)}+z^2 g_{(2)}+z^3 g_{(3)}+O(z^4)\big)_{ij}dx^i dx^j 
\bigg],
\end{eqnarray}
where $x^i=(\theta,\varphi,t)$ and 
\begin{eqnarray}
\label{FG2}
 \big(g_{(0)}+z^2 g_{(2)} \big)_{ij}dx^i dx^j =(L^2-\frac{z^2}{2})(d\theta^2+\sin^2\theta d\varphi^2)-(1+\frac{z^2}{2L^2})dt^2.
\end{eqnarray}
Then the background metric upon which the  dual field theory resides is
$d\sigma^2=g_{(0)ij}dx^i dx^j=-dt^2+L^2 (d\theta^2+\sin^2 \theta d\varphi^2)$, 
which corresponds to a  
static Einstein universe in $(2+1)$ dimensions.

From (\ref{FG1}) one can read the $v.e.v.$ of the holographic stress tensor \cite{deHaro:2000xn}:
\begin{eqnarray}
\label{FG3}
  <\tau_{ij}>=\frac{3L^2}{16\pi G}  g_{(3)ij}=  <\tau_{ij}^{(0)}>+ <\tau_{ij}^{(s)}>,
\end{eqnarray}
being expressed as the sum of a background part plus a matter contribution (which possesses a nontrivial
$\theta-$dependence), 
\begin{eqnarray}
\label{Tij0}
&&
  <\tau_{ij}^{(0)}> dx^idx^j=\frac{1}{16\pi G}\left(\frac{r_H^3}{L^2}+\frac{Q^2+r_H^2}{r_H}\right)
	\left(d\theta^2+\sin^2\theta d\varphi^2+\frac{2}{L^2}dt^2 \right),
	\\
\label{Tijs}
	&&
	  <\tau_{ij}^{(s)}> dx^idx^j=\frac{3}{16\pi G}\frac{1}{L^2} 
	\left( 
	s_{13}(\theta)d\theta^2+f_{23}(\theta)\sin^2\theta d\varphi^2-\frac{1}{L^2}f_{03}(\theta)dt^2 
	\right).
\end{eqnarray}
As expected, this stress-tensor is finite, traceless and covariantly conserved.

\subsubsection{The numerical approach}
 
In our numerical scheme, one starts by choosing a
suitable combination of the  EDT equations together with the matter field(s) equations,
such that the differential equations for the metric and matter function(s) are diagonal
in the second derivatives  with respect to $r$.
Then the radial coordinate $r$ is compactified according to
$r=\frac{x}{1-x},$
with $0\leq x\leq 1$.
All numerical calculations 
are performed by using a professional package
based on the iterative Newton-Raphson method \cite{schoen}. 
The typical  numerical error
for the solutions reported in this work is estimated to be of the order of $10^{-4}$, with smaller   
values for the norm of the $\xi$-field. 
 Further details on the numerical scheme can be found in \cite{Kichakova:2014fta}.

In the numerical calculations, we use units set by the AdS length scale $L$
($e.g.$ we define a scaled radial coordinate, $r\to r/L$).
In practice, this reduces to setting $\Lambda=-3$ in the field equations,
and solving  the EDT equations with 
\begin{eqnarray}
\label{EDT1}
R_{\mu\nu}-\nabla_{(\mu}\xi_{\nu)}=-3 g_{\mu\nu}+2 \alpha^2  
(T_{\mu\nu}-\frac{1}{2}T g_{\mu \nu}),~~
{\rm with}~~\alpha=\frac{\sqrt{4\pi G}}{L}.
\end{eqnarray}
In this approach, $\alpha$ is an input parameter which describes the coupling to gravity;
 the probe limit of the specific models,
which is discussed in Appendix A, has  $\alpha=0$, $i.e.$ a fixed BH geometry.

\section{Static, axially symmetric black holes }
\subsection{ Einstein--scalar field solutions}
\subsubsection{The model}

The simplest static BH solutions with a single Killing vector field
are found in a model with a single complex scalar field, $\Psi \equiv \Phi$,
possessing a Lagrangean density 
 \begin{equation}
\label{action-scalar}
{\cal L}_{m}=
   -\frac{1}{2} g^{\mu\nu}\left( \Phi_{, \, \mu}^* \Phi_{, \, \nu} + \Phi _
{, \, \nu}^* \Phi _{, \, \mu} \right) - U( \left| \Phi \right|) ~,
\end{equation}  
where $U$ is the scalar field potential.
The scalar field is a solution of the Klein-Gordon equation 
\begin{equation}
\label{KG-eqs}
 \frac{1}{\sqrt{-g}} \partial_\mu \big(\sqrt{-g} \partial^\mu\Phi \big)=\frac{\partial U}{\partial\left|\Phi\right|^2} \Phi,
\end{equation}  
with $g_{\mu\nu}$ given by the line-element (\ref{metric}), and possesses a 
 stress-energy tensor  
\begin{eqnarray}
\label{tmunu} 
T_{\mu \nu}  
=
 \Phi_{, \, \mu}^*\Phi_{, \, \nu}
+\Phi_{, \, \nu}^*\Phi_{, \, \mu} 
-g_{\mu\nu} \left[ \frac{1}{2} g^{\alpha\beta} 
\left( \Phi_{, \, \alpha}^*\Phi_{, \, \beta}+
\Phi_{, \, \beta}^*\Phi_{, \, \alpha} \right)+U( \left| \Phi \right|)
\right]
 \ ,
\end{eqnarray}
which enters the EDT equations (\ref{EDT}).

 The scalar field  is complex\footnote{The scalar field ansatz (\ref{s-ans}) is inspired by
 the one employed for AdS spinning Q-balls and boson stars, see $e.g.$ \cite{Radu:2012yx},
 in which case the  phase depends also on time, 
 $\Phi=Z(r,\theta)e^{i (n\varphi-w t)}$.
Then, for $n\neq 0$,
the corresponding configurations would rotate, possessing a single helical 
 Killing vector, without being necessary to ask for $U<0$.
Static axially symmetric solutions are found by taking $w=0$; however, 
this implies 
 turning on a (negative) quartic term in the scalar potential (\ref{pot1}).
 }, with a  phase depending on the azimuthal angle $\varphi$ only, 
\begin{eqnarray}
\label{s-ans}
\Phi=Z(r,\theta)e^{i n\varphi},
\end{eqnarray}
with $n$ a winding number, $ n=\pm 1,\pm 2,\dots$.
This 
allows the energy momentum tensor
to be $\varphi$-independent, even though the scalar field is not; 
hence the ansatz (\ref{s-ans})
is compatible with the symmetries of the  geometry (\ref{metric}).

The value $n=0$ corresponds to the
 spherically symmetric case with a single real scalar field, $\Phi \equiv Z(r)$.
In this limit, the Einstein-scalar field system is known to possess 
BH solutions with AdS asymptotics,
which were extensively discussed in the literature, see $e.g.$ \cite{Hertog:2004dr}.
A necessary condition for the existence of these
  BHs is that scalar field potential $U$
is not strictly positive, which allows for negative energy densities.

\begin{figure}[h!]
\begin{center}
\includegraphics[height=.26\textheight, angle =0]{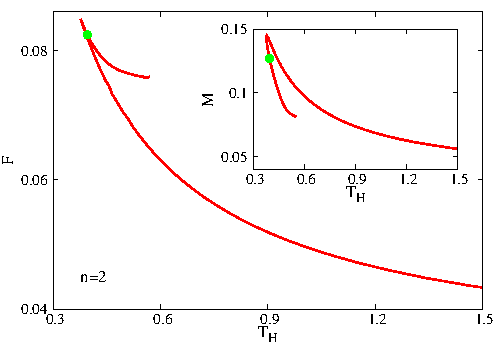} 
\includegraphics[height=.26\textheight, angle =0]{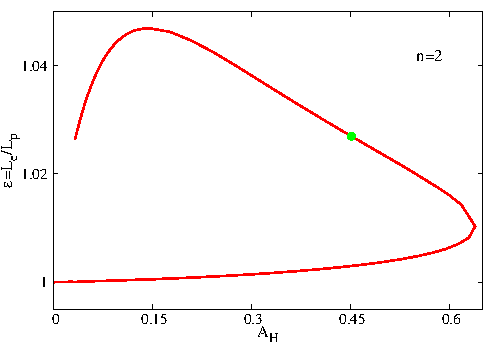} \ \ 
\end{center}
  \vspace{-0.5cm}
\caption{{\it Left panel:} The free energy $F=M-T_H S$ and the mass $M$ 
are shown as functions of the Hawking temperature for a set of 
static axially symmetric black holes in Einstein-scalar field theory
with the input parameters $\lambda=-9$, $\mu=0$ and $\alpha=0.1$.
{\it Right panel:} The horizon excentricity $\epsilon=L_e/L_p$  
is shown as a function of horizon area for the same solutions. 
Note that in all plots in this work, the quantities are expressed in units of $L$, 
the cosmological length scale. }
\label{EKG1}
\end{figure}

\begin{figure}[h!]
\begin{center}
\includegraphics[height=.26\textheight, angle =0]{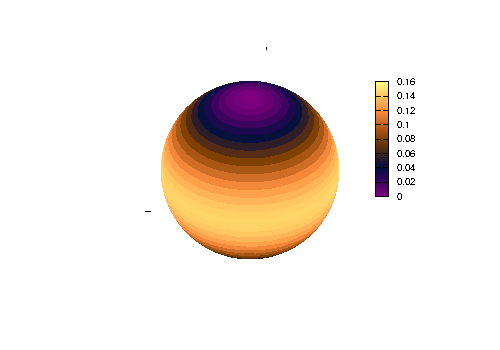} 
\includegraphics[height=.26\textheight, angle =0]{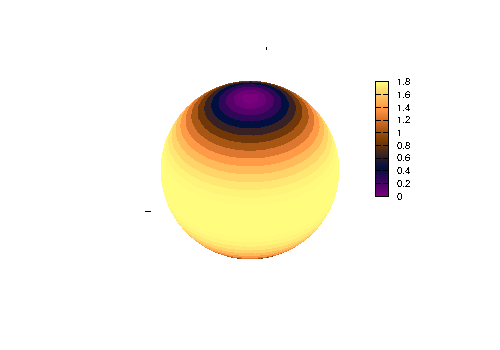} \ \  
\end{center}
 \vspace{-1.5cm}
\caption{The scalar field (left)  and the  energy density (right)  
  on the horizon are shown for an $n=2$ static axially symmetric black hole in Einstein-scalar field theory
marked with a green dot in Figure \ref{EKG1}.}
\label{EKG2}
\end{figure}

However, as we shall argue,
the same mechanism holds for the more general scalar Ansatz (\ref{s-ans}).
In this work we consider a simple potential allowing for $U<0$, 
\begin{eqnarray}
\label{pot1}
U = \mu^2 |\Phi|^2 -\lambda |\Phi|^4,
\end{eqnarray}
where $\lambda$ is a strictly positive parameter and $\mu$ is the scalar field mass.
The constants 
$\mu$ and $L$ fix the behaviour of the scalar field as $r\to \infty$, with
 \begin{eqnarray}
\label{Z-inf}
Z 
\sim 
\frac{c_+(\theta)}{r^{\Delta_+}}+\frac{c_-(\theta)}{r^{\Delta_-}},~{\rm with}~~
\Delta_{\pm}=\frac{3}{2} \left(1\pm \sqrt{1+\frac{4}{9}\mu^2L^2} \right),
\end{eqnarray}
 $c_\pm(\theta)$ being two functions that label the
different boundary conditions.

\begin{figure}[h!]
\begin{center}
\includegraphics[height=.26\textheight, angle =0]{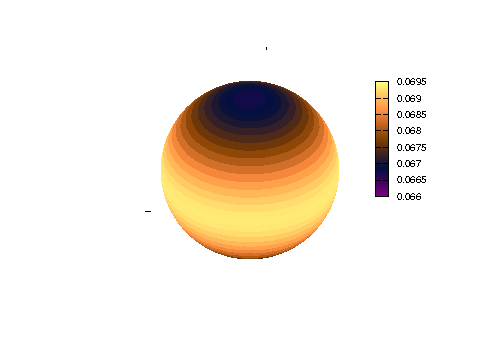} 
\includegraphics[height=.26\textheight, angle =0]{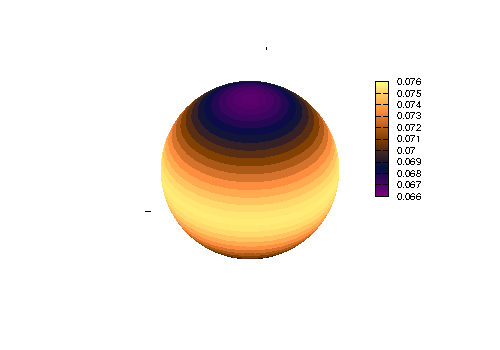} \ \  
\end{center}
 \vspace{-1.5cm}
\caption{The $<\tau_\theta^\theta>$ (left) and $<\tau_\varphi^\varphi>$ (right)
components of the holographic stress tensor are shown for the solution in Fig. \ref{EKG2}. }
\label{EKG3}
\end{figure}

\begin{figure}[h!]
\begin{center}
\includegraphics[height=.26\textheight, angle =0]{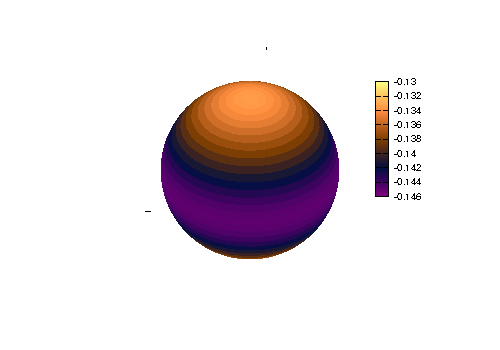} 
\includegraphics[height=.26\textheight, angle =0]{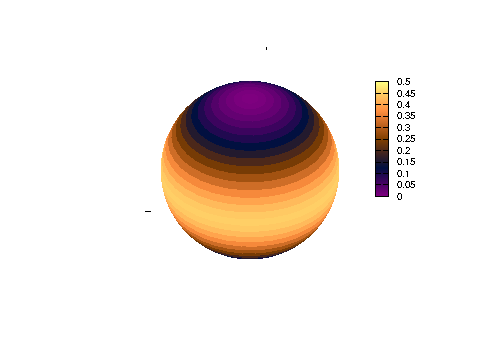} \ \ 
\end{center}
 \vspace{-1.5cm}
\caption{The $<\tau_t^t>$ 
component of the holographic stress tensor  (left) 
and the function $c_+(\theta)$
which enters the asymptotics (\ref{Z-inf}) (right)
are shown for the Einstein-scalar field solution in Fig. \ref{EKG2}. }
\label{EKG4}
\end{figure}

\subsubsection{The results}

The recent work \cite{Kleihaus:2013tba} 
has shown the existence of both  BH and soliton solutions of this model for the  $\Lambda=0$
limiting case,
for the same scalar potential (\ref{pot1}). 

As expected, those configurations
survive in the presence of a negative cosmological constant.
The discussion in this case should start with the observation that
the eq.~(\ref{KG-eqs}) 
possesses solutions already in the probe limit, $i.e.$
when neglecting the backreaction on the spacetime
geometry.
This case is discussed in 
Appendix A1, where we give numerical arguments for the existence of $n>0$
finite mass, regular 
solutions of the (nonlinear) Klein-Gordon
eq.~(\ref{KG-eqs})  in the background of a SAdS BH.
There we notice first the existence of a non-trivial zero horizon
size limit, which describes a static, non-spherically symmetric soliton
in a fixed AdS background.
Second, the solutions do not exist for arbitrarly large SAdS BHs,
with the emergence of a critical horizon radius 
and a secondary branch extending backwards in  $r_H$.

As expected, these solutions survive when taking into account the backreaction
on the spacetime geometry.
In the numerics,  we scale $Z\to Z/L$;
moreover, to simplify the problem, we consider the case $\mu=0$
only.
Then the system possesses an extra scaling symmetry
$r\to r c$, 
$L\to L c$
and $\lambda \to \lambda/c^2$ (with $c>0$ an arbitrary constant)
which can be used to fix the value of $\lambda$ in (\ref{pot1}).
Also, for all solutions, the scalar field is invariant under a reflection 
in the equatorial plane\footnote{
However, the noticed analogy with Q-balls and boson stars
suggests the existence of odd-parity configurations with $Z=0$ at $\theta=\pi/2$, 
while the spacetime geometry is still even parity.}.
Finally, we impose $c_-=0$ in the asymptotic expression (\ref{Z-inf}) 
such that the scalar field decays as $1/r^3$ as $r\to \infty$.

In Figure \ref{EKG1} we give some results of the numerical integration for
a set of $n=2$ solutions with $\Lambda=-3$, $\lambda=-9$, $\mu=0$ and $\alpha=0.1$. 
In the numerics, the control parameter is $r_H$ which fixes the size of the event horizon,
 with $A_H\to 0$ and $T_H\to \infty$
as $r_H\to 0$, a 
  limit which corresponds to a gravitating scalar soliton. 
As $r_H$ increases, both the mass and the event horizon area increase 
while  the temperature decreases.

These BHs cannot be arbitrarly large, and
 we  notice the emergence of a secondary branch of BHs for a critical configuration, 
extending backward in $r_H$, towards $r_H\to 0$.
However, the numerics becomes increasingly challenging on the 2nd branch and
clarifying this limiting behaviour 
remains a task beyond the purposes of this paper. 
Here we only note that the solutions on the first branch minimize 
the free energy 
$F=M-T_H S$
and are thermodynamically favoured.

An interesting feature of the solutions studied\footnote{This result has been found for 
BHs with other values of $n$ and $\mu\neq 0$. However, the deviation from spherical symmetry remains always small.}
 so far is that they always possess an oblate horizon,
$\epsilon=L_e/L_p>1$, as seen in Figure \ref{EKG1} (right) (although the deviation from sphericity is rather small).
Also, both the scalar field $Z$ and the energy density $\rho=-T_t^t$ do  not vanish on the horizon, possessing a strong angular 
dependence\footnote{Note, however, that for a winding number $n=1$, 
the energy density $\rho=-T_t^t$ does not vanish on the symmetry axis  $\theta=0,\pi$.}, as seen in Figure  \ref{EKG2}.
The shape of the scalar field and the energy density in the bulk are rather similar to those 
shown in Appendix A1 for the probe limit,
and we shall not display them here\footnote{Note that the energy density always becomes negative in some
region; however, the total mass is still positive.}.

As seen in Figures  \ref{EKG3}, \ref{EKG4}, the deviation from sphericity of the bulk 
reflects itself in the holographic stress tensor  $<\tau_i^j>$.
One can see that as expected $<\tau_t^t>$ is negative,  while, 
in contrast to the spherically symmetric case,
$<\tau_\theta^\theta>\neq <\tau_\varphi^\varphi>$.  
Moreover, the function $c_+(\theta)$  which enters the asymptotics
(\ref{Z-inf})
of the scalar field possesses also a strong $\theta$-dependence, as seen in  Figure \ref{EKG4} (right).

\subsection{ Einstein--Yang-Mills solutions}
  
\subsubsection{The model}

In order to test the generality of the results above, it is necessary to consider
static axially symmetric BHs with other  matter fields.
For $\Lambda=0$, the first (and still the best known) example
of such solutions has been found in a model with Yang-Mills-SU(2) gauge fields \cite{Kleihaus:1997ic}.
In this case, the matter Lagrangean reads
\begin{eqnarray}
\label{model}
{\cal L}_m= -\frac{1}{2}  
{\rm Tr}
\big \{
F_{\mu\nu}F^{\mu\nu}
 \big \},
\end{eqnarray}
 with the
 field strength tensor
\begin{equation}
F_{\mu\nu} = \partial_\mu A_\nu - \partial_\nu A_\mu + i  [A_\mu, A_\nu] \ ,
\end{equation}
and  the gauge potential
\begin{equation}
A_\mu= \frac{1}{2}\tau_a A_\mu^a,
\end{equation}
 $\tau_a$ being SU(2) matrices.
Variation of   (\ref{model}) with respect to the gauge field $A_\mu$
leads to the YM equations
\begin{eqnarray}
\label{feqA}
  \nabla_\mu F^{\mu\nu}+i [A_\mu, F^{\mu\nu} ]
  =0,
\end{eqnarray}
while the variation with respect to the metric $g_{\mu\nu}$ yields the energy-momentum tensor
of the YM fields
\begin{eqnarray}
\label{Tik}
&T_{\mu\nu}^{(YM)} =
  2 {\rm Tr}
   \big \{
      F_{\mu\alpha} F_{\nu\beta} g^{\alpha\beta}
   -\frac{1}{4} g_{\mu\nu} F_{\alpha\beta} F^{\alpha\beta}
   \big \}
\ .
\end{eqnarray}

As discussed in Appendix A2, similar to the scalar field case discussed above,
 the axially symmetric YM Ansatz 
contains an azimuthal winding number $n$, as seen in eqs.~(\ref{gauge-ansatz})-(\ref{u}).
This Ansatz 
is parametrized by four functions
$H_i(r, \theta)$,
the explicit dependence on $\varphi$
being factorized in the expression of the SU(2) basis,
 such that, in the chosen gauge, $\partial/\partial \varphi$ 
is not a Killing vector of the full EYM system.
However, similar to the scalar case, 
the components of the energy-momentum tensor depend on $(r,\theta)$ only.

A fundamental solution here\footnote{As discussed in the next Section,
it is likely that the Einstein-Maxwell-AdS system
possesses other static BH solutions apart from the RNAdS solution.} is the (embedded-U(1)) Reissner-Nordstr\"om-anti-de Sitter (RNAdS) BH, which has
vanishing potentials,
$H_i=0$ and a net magnetic charge $Q_M=n$.  Apart from that, there are 
also genuine non-Abelian BHs, 
which in the simplest case are spherically symmetric, with $n=1$
and nontrivial potentials $H_2=H_4=w(r)$ while $H_1=H_3=0$.
Different from the (embedded) Abelian case, these configurations may carry a zero net magnetic charge, $Q_M=0$.
An overview of their properties is given in  \cite{Volkov:1998cc}   for  $\Lambda=0$.
An interesting supersymmetric extension of these solutions can be found in  
\cite{Hubscher:2008yz}.

As discussed in 
\cite{Kleihaus:1997ic},
the  $\Lambda=0$ BHs possess static axially symmetric generalizations, which are found by 
taking a value $n>1$ for the azimuthal winding number in the YM ansatz (\ref{gauge-ansatz}).
They possess an event horizon of spherical topology, which, however, is not a round sphere,
and approach  asymptotically a Minkowski spacetime background.
These results have been generalized in 
\cite{Ibadov:2005rb}
by a YM Ansatz containing 
 a further integer, $m>0$, related to the polar angle $\theta$.
In the present formulation of the problem, this integer enters the boundary conditions at infinity
of the gauge potentials, as seen in eqs.~(\ref{bc-odd}) and (\ref{bc-even}).

\subsubsection{The results }

We have found that all known $\Lambda=0$ static axially symmetric EYM BHs possess generalizations 
with AdS asymptotics\footnote{Some properties of the axially symmetric
EYM-AdS BH solutions
were discussed  in a different context in 
\cite{Radu:2004gu,Mann:2006jc} and within  another numerical scheme.
However, the case of configurations with a polar winding number $m>1$ was not 
considered there.}.
Moreover,  new sets of configurations without asymptotically flat
 counterparts do also occur.
The existence of the new solutions can be traced back to the peculiar properties
of the YM system in AdS spacetime. 
In this case,
the confining properties of the AdS geometry (effectively) play the role of a Higgs field, 
leading to a much richer set of possible boundary conditions at infinity, 
as compared to the case $\Lambda=0$.
These  boundary conditions are given in
Appendix A2, where
we provide an overview of the axially symmetric solutions of the YM equations
in a SAdS background, for several values of the integers $(m,n)$ introduced above.
Note that for an (S)AdS background, the boundary conditions satisfied by the YM potentials 
at infinity contain an extra-parameter $w_0$, which fixes the magnetic charge, as seen in eq.~(\ref{Qm}).
This parameter is not fixed apriori,
such that YM solutions with a non-integer magnetic charge 
are allowed
\cite{Winstanley:1998sn,Bjoraker:2000qd,Kichakova:2014fta}.

As expected, these YM configurations survive when taking into account the backreaction
on the spacetime geometry. 
The boundary conditions satisfied by the gauge potentials
$H_i$
at infinity, on the horizon and on the symmetry axis are similar to those used in the probe limit,
as stated in Appendix A2.

\begin{figure}[h!]
\centering
\includegraphics[height=.26\textheight, angle =0]{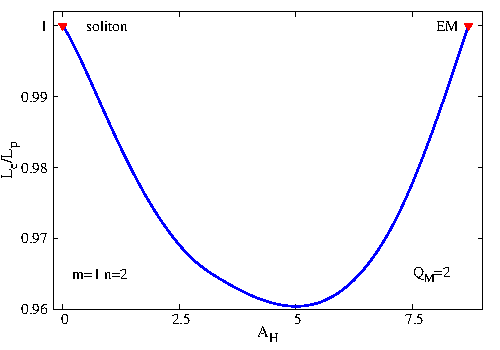} 
\includegraphics[height=.26\textheight, angle =0]{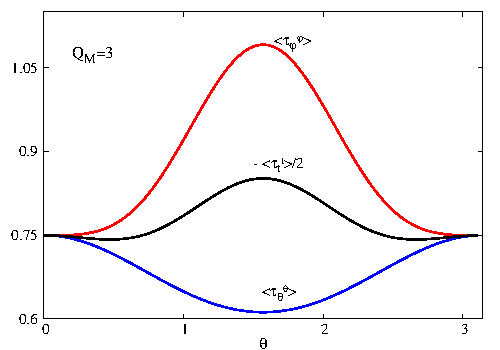} \ \  
\caption{  {\it Left panel:} The ratio $L_e/L_p$ (which gives a measure of the horizon deformation) 
is shown as a function of the horizon area for a set of $m=1,~n=2$  Einstein--Yang-Mills (EYM) 
solutions with $\alpha=0.5$ and magnetic charge $Q_M=2$. 
These configurations interpolate between a horizonless EYM soliton and a critical Reissner-Nordstr\"om-AdS
black hole, see Figure \ref{EYM1} (left).
 {\it Right panel:}  The holographic stress tensor is shown for a typical EYM static
axially symmetric solution with the input parameters $m=1,n=3$, $\alpha=0.5$, $r_H=0.4$
and  $Q_M=3$.
}
\label{EYM3}
\end{figure}

\begin{figure}[h!]
\begin{center}
\includegraphics[height=.26\textheight, angle =0]{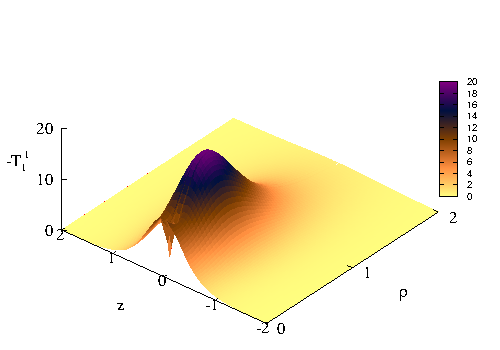} 
\includegraphics[height=.26\textheight, angle =0]{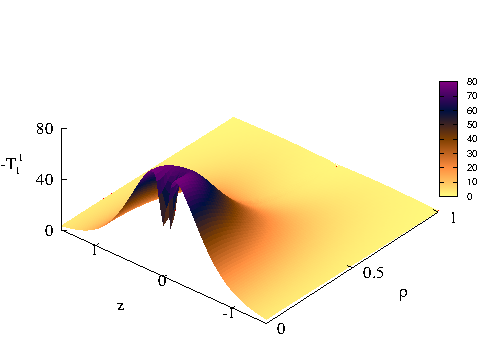} \ \ 
\end{center}
  \vspace{-0.5cm}
\caption{ The energy density is shown for typical $m=1$ (left)
and $m=3$ (right) static axially symmetric solutions 
with vanishing net magnetic charge in Einstein--Yang-Mills theory.
The $m=1$ solution has the input parameters $n=2$, $\alpha=0.15$ and $r_H=0.05$, while the one with
$m=3$ has $n=1$,   $\alpha=0.1$ and $r_H=0.05$. 
The axes here are $\rho= \bar r \sin \theta$, $z=\bar r \cos \theta$, 
with $\bar r=\sqrt{r^2+r_H^2}$.
}
\label{EYM1s}
\end{figure}

\begin{figure}[h!]
\centering
\includegraphics[height=.26\textheight, angle =0]{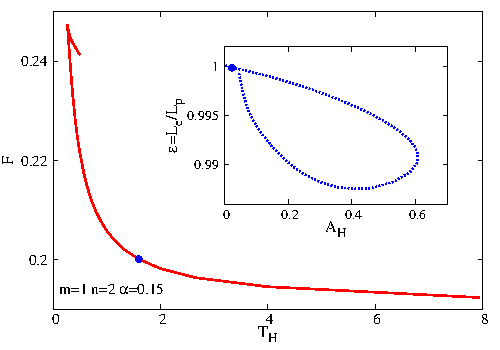} 
\includegraphics[height=.26\textheight, angle =0]{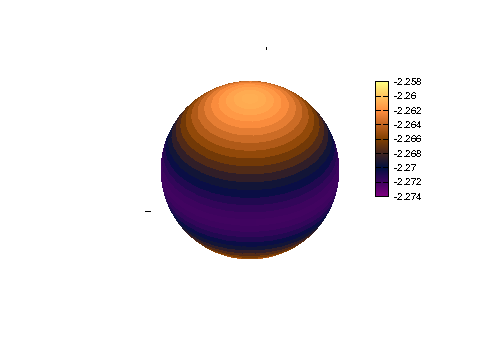} \ \  
\caption{  {\it Left panel:} 
The free energy $F=M-T_H S$ is shown as a function of the Hawking temperature for a set of
 Einstein--Yang-Mills static axially symmetric black holes   with a vanishing net magnetic charge,
$Q_M=0$.
The inset shows the ratio $L_e/L_p$ (which gives a measure of the horizon deformation) 
  as a function of horizon area for the same set of black holes. 
 {\it Right panel:}  The $<\tau_t^t>$ component of the holographic stress tensor is shown for a
 solution marked with a dot on the left panel.
}
\label{EYM3s}
\end{figure}

An interesting feature shared by all EYM axially symmetric BHs studied so far, is that 
 in contrast to the solutions with a scalar field, 
they possess a prolate horizon,
$\epsilon =L_e/L_p<1$,
 although the horizon's deviation from sphericity remains small, as seen in Figure \ref{EYM3} (left).
However, as shown in Figures \ref{EYM3}, \ref{EYM3s} (right panels), 
the holographic stress tensor exhibits a strong
$\theta$-dependence, with $<\tau_t^t>$ always strictly negative.
Also, the energy density shows a strong angular dependence, as seen in Figure \ref{EYM1s};
in particular its value on the horizon varies with $\theta$. 

%
%

Concerning further properties of the solutions,
the results 
can be summarized as follows\footnote{
Within our formulation,
the numerical problem possesses five input parameters: $(r_H,\alpha; m,n)$ and $w_0$ (which fixes the magnetic charge).
The emerging overall picture is rather complicated 
and we did not attempt to explore in a systematic way the parameter space of all solutions.
}.
First, we have found that all YM configurations in a fixed SAdS background
possess gravitating generalizations.
The backreaction is taken into account by slowly increasing the 
value of the parameter $\alpha=\sqrt{4\pi G}/{L}$. 
As the coupling constant $\alpha$ increases, the
spacetime geometry is more and more deformed. 
However, $\alpha$
cannot be arbitrarily large. The solutions stop to exist at
a critical value\footnote{It is interesting to note that a
similar behavior is found for the (asymptotically flat) 
solutions of the EYM-Higgs system, the role of 
$\Lambda$ there being played by the Higgs field \cite{Hartmann:2001ic}.}
which, for a given $Q_M$, depends on both integers $(m,n)$
and on the size of the horizon, as specified by $r_H$.
There, a second branch of solutions emerges, which extends backward towards $\alpha\to 0$.
Indeed, this limit can be approached in two ways: 
(i) for the Newton constant $G \to 0$, or,
(ii) for $L\to \infty$ (or, equivalently, $\Lambda \to 0$). In the former case
one recovers the YM solutions in a fixed  SAdS
background, while the latter case (with $Q_M=0$) corresponds to the known EYM
hairy BH in the asymptotically flat spacetime 
\cite{Kleihaus:1997ic,Ibadov:2005rb}.
In contrast, the configurations with $Q_M \neq 0$
do not possess a well-defined asymptotically flat limit.

Second, all configurations possess a nontrivial 
 horizonless, particle-like limit,
which is approached as the horizon size shrinks to zero.
The corresponding solitonic solutions have been extensively
 studied in 
 \cite{Kichakova:2014fta}
(and their asymptotically flat counterparts
in \cite{Kleihaus:1996vi,Ibadov:2005rb}).
 The most interesting feature found there 
 is the existence  for $m>1$  of balanced, regular composite configurations, with several distinct components.
 Our results show that, as expected, all these solutions survive
 when including a  horizon at their center of symmetry.

\newpage
%
\begin{figure}[h!]
\begin{center}
\includegraphics[height=.26\textheight, angle =0]{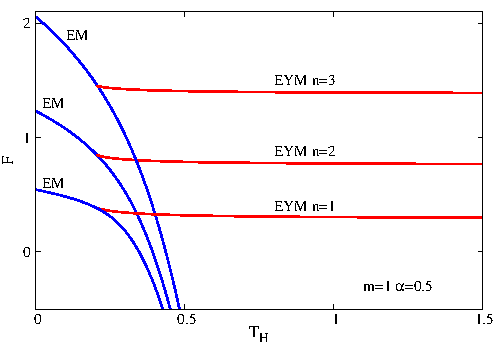} 
\includegraphics[height=.26\textheight, angle =0]{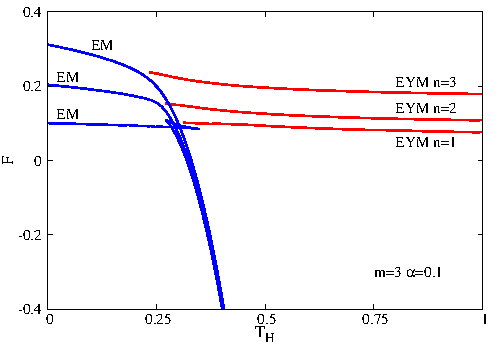} \ \ 
\end{center}
  \vspace{-0.5cm}
\caption{
The free energy $F=M-T_H S$ is shown as a function of the Hawking temperature for several sets of 
static axially symmetric black holes in Einstein--Yang-Mills (EYM) theory,
together with the Einstein-Maxwell (EM) solutions with the same magnetic charge.
These solutions possess a net magnetic charge $Q_M=n$.
 }
\label{EYM1}
\end{figure}
%

In our study we have paid special attention to
(possibly) the  most interesting case, corresponding to
 configurations 
whose YM far field asymptotics describe a gauge transformed charge$-n$ Abelian (multi)monopole.
Then, to leading order, the far field asymptotics of the
EYM solutions are identical to those of an embedded RNAdS BH with a magnetic charge \footnote{Note that in the absence of a Higgs field, 
such configurations do not have asymptotically flat counterparts.
For $\Lambda<0$,
the AdS geometry supplies the attractive force needed to balance the repulsive force of Yang-Mills gauge interactions.
This can seen $e.g.$ from the study of the exact $m=n=1$ 
unit magnetic charge solution of the YM equations found in \cite{BoutalebJoutei:1979va}.
} $Q_M=n$.

These solutions have a number of interesting thermodynamical features,
whose systematic study is, however, beyond the purposes of this work.
Here we mention only that
the existing results suggest that the picture found in \cite{Kichakova:2015lza}
for $m=n=1$ spherically symmetric solutions can be generalized to the axially symmetric case.
That is, for a given $\alpha$, the branches of solutions possessing a net integer magnetic charge 
 bifurcate from some critical RNAdS configuration with the same $(Q_M,\alpha)$. 
Also, for all configurations, 
 there exists an (embedded-Abelian) charge$-n$ RNAdS solution which is  
thermodynamically favoured over the non-Abelian ones, as seen in Figure \ref{EYM1}.

We have  studied also several sets of 
 $m=1$ and $m=3$ solutions  with a vanishing net magnetic flux $Q_M=0$.
(Note, however, that the bulk magnetic charge density is nonzero.)
The simplest solutions here have $m=1,n=1$, being spherically symmetric 
\cite{Winstanley:1998sn,Bjoraker:2000qd}.
(These configurations can be considered as the natural counterparts of the $\Lambda=0$
EYM BHs \cite{Volkov:1998cc}.) 
Axially symmetric generalizations of the $m=1$ BHs
 are found by taking  $n \geq 2$ in the YM ansatz (\ref{gauge-ansatz}).
The configurations with $m=3$
do not possess a spherically symmetric limit.
Some of the basic properties of these solutions are different 
as compared to the case with a net magnetic charge $Q_M\neq 0$ above. 
For example, they do not emerge as
perturbations  of an embedded Abelian solution.
Then the geometry  of the (zero charge) SAdS BH is approached (to leading orders) 
in the asymptotic region only.
The thermodynamics of the solutions is also different as compared to the magnetically charged case,
our results suggesting that the picture found
in the recent work  \cite{Kichakova:2015lza}  for  $m=1,n=1$ BHs
remains valid in the axially symmetric case.
For example,
in a free energy-temperature diagram, one finds two branches of solutions, 
which form a cusp for some minimal value of $T_H$, as seen in Figure \ref{EYM3s} (left panel).  

Finally, let us also mention the existence of configurations with 
a non-integer (and non-vanishing) magnetic charge.
Although we did not attempt to investigate these solutions in a systematic way,
we confirm that such EYM BHs also possess a prolate event horizon.

\section{Further remarks.
 Static axially symmetric black holes in Einstein-Maxwell-AdS theory?
}

The main purpose of this paper was to
 show the existence of asymptotically AdS$_4$ static BHs
which are axially symmetric only. 
As explicit examples, we have considered first the case of 
Einstein-(complex) scalar field theory, followed by
a study of static axially symmetric BHs in 
Einstein--Yang-Mills theory.

In both cases the mechanism which allowed for 
the existence of such BHs was 
the {\it symmetry non-inheritance} of the matter fields \cite{Smolic:2015txa}.
That is, the matter fields  
possess a dependence 
on the azimuthal angle $\varphi$, which, however, does not manifest at the 
level of the energy-momentum tensor. 
Similar solutions should exist in various other models with this feature,
$e.g.$ Einstein-Skyrme or Einstein--Yang-Mills--Higgs.

There are a number of common features for both types of solutions discussed in this work.
For example, they possess a smooth particle-like 
horizonless limit;
as such,  both classes of BHs follow the paradigm of {\it `event
horizons inside classical lumps'} \cite{Kastor:1992qy}.
They can also be considered as {\it `bound states of an
ordinary black hole and a soliton'}
\cite{Ashtekar:2004cn}.
Moreover, the size of such BHs
cannot be arbitrarily large, with the occurrence of
a secondary branch of solutions for a critical configuration which maximizes the horizon area.
We also note that some of their basic features can be derived based on the results
obtained in the probe limit ($i.e.$ matter field(s) in a fixed BH background).

Perhaps the most unexpected feature of the solutions in this work 
is that the excentricity of the horizon (as measured by $\epsilon=L_e/L_p$),
changes from oblate to prolate when considering BHs in Einstein--scalar field
theory and EYM theory.
Thus, the internal (matter field(s)) interactions lead to a different shape of  the horizon.
A better understanding of this feature will require
a study of static axially symmetric solutions with  a more general matter content.
In particular, it will be interesting to consider systems containing gauged scalars\footnote{For partial results
in this direction, see $e.g.$ \cite{Kichakova:2013sza}.
}.
Let us also mention that this kind of configurations 
(including those in this work) should allow for 
spinning generalizations,
 which would possess a 
single $helical$ Killing vector\footnote{
Remarkably, as shown in the recent work \cite{Dias:2015rxy},
such BHs exist already in the vacuum case.
}.

Another possible direction would be the construction 
of static solitons and BHs with AdS asymptotics, which possess discrete
symmetries  only.
The existence of such solutions is suggested by the 
results in the literature obtained for a Minkowski spacetime background,
the case of Einstein-Skyrme theory being possibly the simplest example\footnote{Again, this geometric feature can be viewed as an 
imprint  of symmetry non-inheritance  of the matter fields.} 
\cite{Ioannidou:2006mg}.(A different example is given in
\cite{Kleihaus:2004wu}.)

One interesting question to ask here is whether the 
symmetry non-inheritance  of the matter fields
is the only mechanism leading to static BHs which
are axially symmetric only.
As argued below, the answer to this question is likely to be negative.
For this purpose,
in the remainder of this Section, we now
discuss the possible existence 
of static axially symmetric BH solutions in Einstein-Maxwell (EM) theory. 
Different from the cases in Section 3, the
matter field ($i.e.$ the  U(1) potential)
 inherits the spacetime symmetries.
Then the
possible existence of such EM configurations would be anchored
this time 
in the confining properties of the AdS spacetime.
 
The starting point here 
is the study of a more general asymptotical behavior of the YM fields 
in an AdS background as compared to
the one considered in Appendix A2.
Another hint comes from 
 the results in the recent paper \cite{Herdeiro:2015vaa}.
There it was shown 
that the Maxwell equations in a fixed
AdS background possess
everywhere regular solutions, with finite energy, for every electric multipole moment  
\textit{except} for the monopole. 
Let us briefly review this result.
It is well-known that a static axially symmetric charge distribution in a flat spacetime background
possesses a multipolar expansion for the electrostatic potential of the form
\begin{eqnarray}
\label{s1}
V(r,\theta)=\sum_{\ell \geq 0}R_{\ell}(r) \mathcal{P}_\ell(\cos \theta), 
 ~{\rm with}~~R_{\ell}(r) =c_1 R_{\ell}^{(1)}(r)+c_2 R_{\ell}^{(2)}(r),~
{\rm and}~
R_{\ell}^{(1)}(r)=r^\ell,~~R_{\ell}^{(2)}(r)=\frac{1}{r^{\ell+1}},~
\end{eqnarray}
(with $c_1$, $c_2$ arbitrary constants and 
$\mathcal{P}_\ell$ the Legendre polynomial of degree $\ell$)
such that any solution diverges either at $r=0$ or for $r\to \infty$.
However, as explicitly shown in \cite{Herdeiro:2015vaa},
the situation is different for an AdS background, which regularizes the (far field)
divergence in $R_{\ell}^{(1)}(r)$ (with $\ell \geq 1$),
which now approaches a constant value
as $r\to \infty$.  
 As usual, the existence of such solutions in the probe limit
 is taken as a strong indication that the full Einstein-Maxwell
system possess nontrivial gravitating configurations.
Indeed,  Ref.
\cite{Herdeiro:2015vaa}
has computed the
first order perturbation induced  in the AdS geometry by a regular electric dipole
and found an exact solution  
showing that all metric functions remain smooth.

However, some of the results in 
\cite{Herdeiro:2015vaa}
hold also when replacing the 
AdS background 
with a SAdS BH.
Starting again with a general axially symmetric electrostatic potential
$V(r,\theta)=\sum_{\ell \geq 0}R_{\ell}(r) \mathcal{P}_\ell(\cos \theta)$,
we show in
the  Appendix A3 that
for $\ell\geq 1$
 and given $r_H\geq 0$, the Maxwell equations
possess a solution which is smooth, with finite energy.
Unfortunately, this solution cannot be found in closed form
(except for $r_H=0$ $i.e.$ a globally AdS background); however, it can
easily be constructed numerically.

We expect these configurations to survive when including the backreaction
on the spacetime geometry (as seen in this work, this was the case for both
scalar and YM fields). 
Thus it is natural to conjecture\footnote{Following Ref.~\cite{Herdeiro:2015vaa},
one can 
approach this problem perturbatively,
the perturbative parameter being the magnitude of the electrostatic
potential at infinity.
In the absence of an exact solution in the probe limit, this reduces 
to solving numerically a set of ordinary differential equations with suitable boundary conditions.
A nonperturbative approach appears also possible,
requiring an adjustment of the scheme described in Section 2.}
 that  {\it `the Reissner-Nordstr\"om-AdS solution is not the
unique static BH in Einstein-Maxwell theory with a negative
cosmological constant, with the existence of a different type of
AdS BHs 
which are static and axially symmetric only'}.

 \vspace*{0.5cm}
\noindent{\textbf{~~~Acknowledgements.--~}}
We would like to thank C.~Herdeiro 
and B.~Kleihaus for relevant discussions.
We gratefully acknowledge support by the DFG Research
Training Group 1620 {\it `Models of Gravity'},
by the Alexander von Humboldt Foundation in
the framework of the Institutes Linkage Programme,
and by FP7, Marie Curie Actions, People, 
International Research Staff Exchange Scheme (IRSES-606096).
E.R.~gratefully acknowledges support from the FCT-IF programme.

\appendix


\section{The probe limit}
\label{appendixa}

In this Appendix we consider static solutions for several different models in a SAdS background given by
\begin{eqnarray}
\label{SADS}
ds^2=\frac{d \bar r^2}{N(\bar r)}+\bar r^2(d\theta^2+\sin^2\theta d\varphi^2)-N(\bar r) dt^2,
\end{eqnarray} 
where 
\begin{eqnarray}
\label{NBH}
N(\bar r)= \left(1-\frac{r_H}{\bar r} \right)\left(1+\frac{\bar r^2}{L^2}+\frac{r_H(r_H+\bar r)}{L^2} \right).
\end{eqnarray}
This geometry possesses a (non-extremal) event horizon at $\bar r=r_H>0$,
the BH mass being $M=\frac{r_H}{2}(1+\frac{r^2_H}{L^2})$.

The energy density of a given field configuration, $\rho$, as measured by a static observer with 4-velocity $U^\mu\propto \delta^\mu_t$, 
is $\rho=-T_t^t$. 
The corresponding total mass-energy of the solutions is
\begin{eqnarray}
M=-\int d^3 x  \sqrt{-g}\, T_t^t=-2\pi \int_{r_H}^\infty d\bar r  \int_0^\pi d\theta~ \sin \theta~\bar r^2  T_t^t.
\end{eqnarray}

We solve  the matter field equations in the region
outside the event horizon only, $\bar r\geq r_H$. 
In our numerical approach for both scalar and Yang-Mills fields,  a new radial
coordinate is introduced, $r=\sqrt{\bar r^2-r_H^2}$, with $0\leq r<\infty$,
such that the horizon is located at $r=0$.
This leads to a simple near horizon expression of the matter fields and, subsequently,
to a simple set of boundary conditions there, which are Neumann or Dirichlet only.

\subsection{The scalar field in a SAdS background}

The matter Lagrangean and the scalar field ansatz
are given in Section 3.1, by the relations
 (\ref{action-scalar}) 
and (\ref{s-ans}), respectively,
with $\Phi=Z(r,\theta) e^{in \varphi}$ and $n=1,2,\dots$. 
The scalar amplitude $Z$ satisfies 
the following boundary conditions:
\begin{eqnarray}
 \partial_r Z|_{r=0}=0,~~Z|_{r=\infty}=0,~~Z|_{\theta=0,\pi}=0.
\end{eqnarray}

\begin{figure}[h!]
\centering
\includegraphics[height=2.15in]{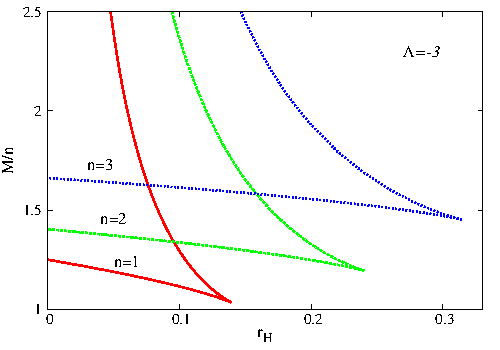}  
\caption{  The mass  (divided by the winding number $n$) of the  static
axially symmetric scalar bound states in a Schwarzschild-AdS background
is shown as a function of the event horizon radius.
}
\label{KG1}
\end{figure} 
%
\begin{figure}[h!]
\centering
\includegraphics[height=.26\textheight, angle =0]{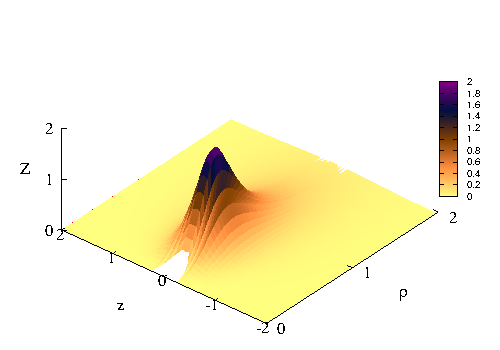} 
\includegraphics[height=.26\textheight, angle =0]{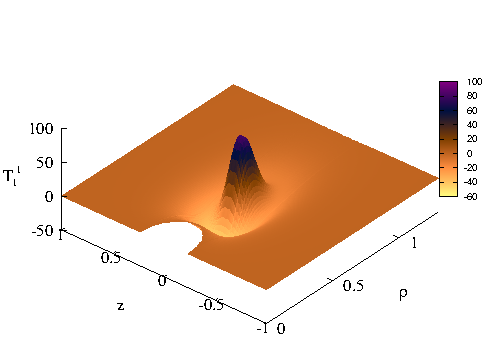} \ \ 
\caption{  The profile of the scalar field amplitude $Z$
and the $T_t^t$ component of the energy-momentum 
tensor are shown for an $n=2$ solution of the non-linear Klein-Gordon equation
 in a Schwarzschild-AdS background (\ref{SADS}) 
with $r_H=0.23$, $L=1$.
The axes here are $\rho= \bar r \sin \theta$, $z=\bar r \cos \theta$.
}
\label{KG2}
\end{figure} 

The results of the numerical integration of the nonlinear KG equation (\ref{KG-eqs})
 are shown in Figure \ref{KG1}. (All results in this Subsection have the parameters in the scalar potential $\mu=0$, $\lambda=10$.)
Here we exhibit the total mass $vs.$ the horizon size as given by the parameter
 $r_H$, for several values of the winding number $n$.
One can notice the existence of two branches of solutions,
which form a cusp for some critical $r_H$, whose value  increases with $n$.
The first branch of solutions emerges from the solitonic configurations
(which have $r_H=0$, $i.e.$ an AdS background).
The total mass-energy of the second branch of solutions appears to diverge as $r_H\to 0$. 
A typical solution is shown in Figure \ref{KG2},
where we display both the scalar amplitude $Z$ and the energy density.

\subsection{The YM fields in a SAdS background}

The YM model has been introduced in Section 3.2, see the relations (\ref{model})-(\ref{Tik}).
The corresponding axially symmetric Ansatz is more complicated as compared to the scalar field case,
and we shall briefly discuss it here.
Following the work \cite{Kleihaus:1997ic}, we choose a gauge potential $A$
containing  four gauge field functions $H_i$ which depend on $r$ and $\theta$ only;
however, it depends also  on the azimuthal coordinate $\varphi$,
with 
 \begin{eqnarray}
\label{gauge-ansatz}
 A_\mu dx^\mu=
\left( \frac{H_1}{r} dr + (1-H_2)d\theta\right)\frac{u_\vphi^{(n)}}{2 }
- n \sin\theta \left( H_3\frac{u_r^{(n)}}{2 }
                     + (1-H_4)\frac{u_\theta^{(n)}}{2 }\right) d\vphi.
\end{eqnarray}
$u_a^{(n )}$ are SU(2) matrices which factorize the dependence on the azimuthal coordinate $\varphi$,
with
\begin{eqnarray}
\nonumber
u_r^{(n )}&=&\sin  \theta (\cos n\varphi~\tau_x+\sin n\varphi~\tau_y)+\cos \theta ~\tau_z,
\\
\label{u}
u_\theta^{(n )}&=&\cos \theta (\cos n\varphi~\tau_x+\sin n\varphi~\tau_y)-\sin \theta ~\tau_z,
\\
\nonumber
u_\varphi^{(n)}&=&-\sin n\varphi~\tau_x+\cos n\varphi~\tau_y ,
\end{eqnarray}
where 
 $ \tau_x, \tau_y, \tau_z $ are the Pauli matrices.
The positive integer $n$ represents the  azimuthal winding number of the solutions.
This ansatz is axially symmetric in the sense that a rotation around the symmetry axis
can be compensated by a gauge rotation,
$\partial_\varphi A_\mu=D_\mu\bar \Psi$ 
\cite{Manton:1977ht,Rebbi:1980yi,Forgacs:1980zs},
with $\bar\Psi$ being a Lie-algebra valued gauge function\footnote{ For the Ansatz (\ref{gauge-ansatz}),
$\bar\Psi=n \cos\theta u_r^{(n)}/2-n\sin\theta u_\theta^{(n)}/2.$}
and $D_\mu$ the gauge covariant derivative.
%
The corresponding expression of the field strength tensor can be found $e.g.$ in Ref.~\cite{Kichakova:2014fta}.
(Note that $F_{\mu\nu}$ also  depends on $\varphi$.)

As usually, in order to fix the residual $U(1)$ gauge invariance of the Ansatz (\ref{gauge-ansatz}),
 we impose the gauge fixing condition
 $r \partial_r H_1- \partial_\theta H_2 = 0$ \cite{Kleihaus:1997ic}.

\newpage

\begin{figure}[h!]
\begin{center}
\includegraphics[height=.26\textheight, angle =0]{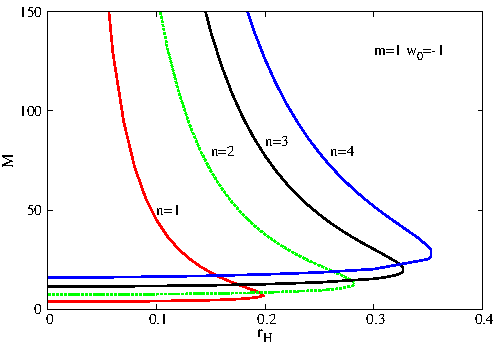} 
\includegraphics[height=.26\textheight, angle =0]{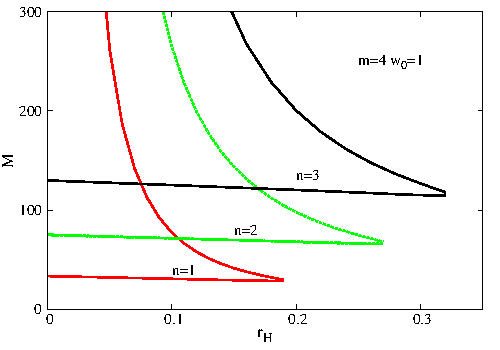} \ \ 
\end{center}
  \vspace{-0.5cm}
\caption{ The mass-energy $M$ is shown as a function of the horizon radius $r_H$ for different sets of YM solutions
in a Schwarzschild-AdS background. }
\label{YM2}
\end{figure}
%

The  magnetic potentials $H_i(r, \theta)$ 
 satisfy a suitable set of boundary conditions at the horizon, at infinity
and on the symmetry axis imposed by finite energy, regularity and symmetry requirements.
At the horizon ($r=0$) we impose 
\begin{eqnarray}
\label{origin}
 H_{1}|_{r=0}=\partial_r H_{2}|_{r=0}= 
\partial_r H_{3}|_{r=0}=\partial_r H_{4}|_{r=0}= 0,
\end{eqnarray} 
while the boundary conditions on the symmetry axis are
\begin{eqnarray}
H_1|_{\theta=0,\pi }=H_3|_{\theta=0,\pi }=0,~~
\partial_\theta H_2|_{\theta=0,\pi }
=\partial_\theta H_4|_{\theta=0,\pi }
=0.
\end{eqnarray} 
At infinity, the gauge potentials are required to satisfy a set of boundary
conditions originally proposed in  \cite{Kichakova:2014fta}.
 The data there contains a new positive integer, $m$,
 which is interpreted as a polar winding number solutions. 
Namely, for odd values of $m$, one imposes 
\begin{eqnarray}
&&
\nonumber
H_1|_{r=\infty}=0,~~H_2|_{r=\infty}=1-m+w_0,
\\
\label{bc-odd}
&&
H_3|_{r=\infty}=\frac{\cos \theta}{\sin \theta} \bigg(\cos ((m-1)\theta)-1\bigg)+w_0\sin((m-1) \theta),
\\
\nonumber
&&
H_4|_{r=\infty}=-\frac{\cos \theta}{\sin \theta} \sin ((m-1)\theta) +w_0\cos((m-1) \theta),
\end{eqnarray} 
while
for even values of $m$ one imposes instead
 \begin{eqnarray}
\nonumber
&&H_1|_{r=\infty}=0,~~H_2|_{r=\infty}=1-m-w_0,~~\\
\label{bc-even}
&&H_3|_{r=\infty}=w_0\frac{\cos((m-1)\theta)-\cos\theta}{\sin\theta},~~\\
\nonumber
&&H_4|_{r=\infty}=1-w_0-w_0\frac{\sin((m-1)\theta)}{\sin\theta},
 \end{eqnarray} 
The parameter $w_0$ which enters the above relations in an arbitrary constant
 which fixes the magnetic charge of the solutions \cite{Kichakova:2014fta}.
One finds 
\begin{eqnarray}
\label{Qm}
Q_M=n|1-w_0^2|~~{\rm for~odd~} m,~~ {\rm and}~~ Q_M=\frac{mn}{2}|w_0(1-w_0)|~~{\rm for~even~} m.
\end{eqnarray} 

 Ref. \cite{Kichakova:2014fta} has given an extensive discussion of the
YM solutions in a fixed AdS background, within this general framework.
As expected, all those solutions can be generalized by replacing 
the regular origin with a BH horizon.
First, we have solved the YM equations for a large set of $(m,n)$
integers and several (fixed) values of the constant $w_0$ in (\ref{bc-odd})-(\ref{bc-even}).
Some results in this case are shown in Figure \ref{YM2}, where 
 we exhibit the total mass of the solutions as functions of the size of the BH
as given by the
 horizon radius $r_H$.
One can notice the existence of two branches of solutions.
The lower branch starts with the solution in a fixed AdS background and merges
for a critical $r_H$ with a secondary branch.
This secondary branch extends backward in $r_H$, 
the mass  increasing with decreasing the horizon radius. 
Also, we have found that the maximal value of the horizon 
radius increases with
decreasing $|w_0|$. 

We have also  studied, for given sets of $(m,n)$,
solutions in a SAdS background with fixed horizon radius
and a varying $w_0$ in the asymptotic boundary conditions (\ref{bc-odd})-(\ref{bc-even}) 
($i.e.$ the magnetic charge).
A general feature here is that solutions exist for a limited range of $w_0$ only,
and possess a rather complicated branch structure.
Thus, for a given horizon size,
one cannot find YM configurations with an arbitrarily large 
(non-Abelian) magnetic charge.

\subsection{Maxwell field multipoles in SAdS background}

The  Lagrangean for a Maxwell field reads
\begin{eqnarray}
{\cal L}_m=-\frac{1}{4}F_{\mu \nu}F^{\mu\nu}, 
\end{eqnarray}
where $F_{\mu \nu}=\partial_\mu A_\nu-\partial_\nu A_\mu$ is the $U(1)$ field strength.
The four-potential $A_\mu$ satisfies the Maxwell equations
\begin{eqnarray}
\label{Maxwell-eqs}
\nabla_\mu F^{\mu \nu}=0,
\end{eqnarray}
with a line element given by (\ref{SADS}). (Note that in order to simplify the relations we take $\bar r\to r$ in what follows.)
The corresponding energy-momentum tensor is
\begin{eqnarray}
T_{\mu\nu}=F_{\mu \alpha}F_{\nu\beta}g^{\alpha \beta}-\frac{1}{4}g_{\mu\nu}F^2~.
\end{eqnarray}

Following 
\cite{Herdeiro:2015vaa},
we consider first a purely electric  $U(1)$ potential $A$ which possesses axial  symmetry only
\begin{eqnarray}
\label{ep}
A_\mu dx^\mu=V(r,\theta)dt = R_\ell(r) \mathcal{P}_\ell(\cos \theta) dt\ , 
\end{eqnarray}
where 
 $\mathcal{P}_\ell$ is a Legendre polynomial of degree $\ell$, with $\ell=0,1,\dots$.

  From (\ref{Maxwell-eqs}) it follows that 
the radial function $R_\ell(r)$
 is a solution of the equation
 \begin{eqnarray} 
 \label{radial}
\frac{d}{dr} \left(r^2\frac{dR_\ell(r)}{dr}\right)=\frac{\ell(\ell+1)}{N(r)}R_\ell,
\end{eqnarray}
with $N(r)$ given by (\ref{NBH}).
Unfortunately, this equation cannot be solved in closed form\footnote{
For $r_H=0$, the solution of (\ref{radial}) reads \cite{Herdeiro:2015vaa}
$
R_\ell(r)=
\displaystyle{\frac{\Gamma(\frac{1+\ell}{2})\Gamma(\frac{3+\ell}{2})}{\sqrt{\pi}\Gamma(\frac{3}{2}+\ell)}
 \frac{r^{\ell}}{L^\ell}~{}_2F_1\left(\frac{1+\ell}{2} , \frac{ \ell}{2}   ;  \frac{3}{2}+\ell  ;  - \frac{r^2}{L^2}\right)} .
$
} for a SAdS background, except for $\ell=0$, with $R_0=c_0-c_1/r$.
However, ( \ref{radial}) can easily be solved numerically for any $\ell\geq 1$; one can also construct an approximate
solution at the limits of the $r-$interval.
The radial function vanishes on the horizon; the solution there
can be
written as a power series in $(r-r_H)$, the first terms being 
\begin{eqnarray}
R_{\ell}(r)=r_1(r-r_H)+\frac{r_1\left ((\ell-1)(\ell+2)-\frac{6r_H^2}{L^2} \right)}{2r_H(1+\frac{3r_H^2}{L^2})}(r-r_H)^2+O(r-r_H)^3,
\end{eqnarray}
%
%
%
\begin{figure}[h!]
\begin{center}
\includegraphics[height=.26\textheight, angle =0]{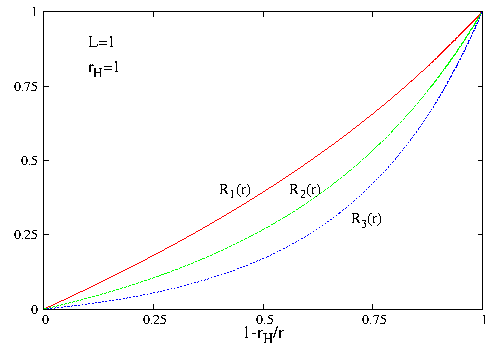} 
\includegraphics[height=.26\textheight, angle =0]{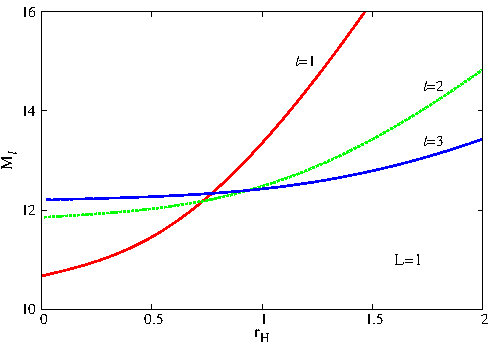} \ \ 
\end{center}
  \vspace{-0.5cm}
\caption{{\it Left panel:}  The radial function $R_\ell$ (with $\ell=1,2,3$)
 is shown in terms of a compactified coordinate
$1-r_H/r$ for Maxwell field solutions in a fixed Schwarzschild-AdS background.
{\it Right panel:} The  total mass-energy for families of $\ell=1,2,3$ solutions
is shown as a function of the event horizon radius. 
In both plots, we consider a normalization $R_\ell\to 1$ asymptotically.}
\label{Maxwell1}
\end{figure}
 \\
where $r_1$ is a parameter which results from the numerics\footnote{
Note that for $r_H=0$, one finds 
$R_\ell(r)\to \frac{ \Gamma\left(\frac{1+\ell}{2}\right) 
\Gamma\left(\frac{3+\ell}{2}\right)}{\sqrt{\pi} \Gamma\left(1+\frac{\ell}{2}\right)}  \left(\frac{r}{L}\right)^\ell$, as $r\to 0$
\cite{Herdeiro:2015vaa}.}.

As $r\to \infty$, the solution reads
\begin{eqnarray}
R_{\ell}(r)=1-c_1^{(\ell)}\frac{L}{r}+\frac{1}{2}\ell(\ell+1)\frac{L^2}{r^2}+\dots,
\end{eqnarray}
where we normalized it such that $R_{\ell}(r)\to 1$ asymptotically.
For $r_H=0$, one finds
$
c_1^{(\ell)}= \frac{2 \Gamma\left(\frac{1+\ell}{2}\right) \Gamma\left(\frac{3+\ell}{2}\right)}{ \Gamma\left(1+\frac{\ell}{2}\right) \Gamma\left(\frac{\ell}{2}\right)}$;
in a BH background, its value is found numerically.

In Figure~\ref{Maxwell1} (left) 
we exhibit the radial function $R_{\ell}$
for a SAdS background with a fixed horizon radius $r_H=1$ and $\ell=1,2,3$.
The dependence of the total mass-energy of the $\ell=1,2,3$ solutions as a function 
of the event horizon radius is shown in Figure \ref{Maxwell1} (right).
Note that in both plots we take an AdS length scale $L=1$.

Finally, let us mention that given the electric-magnetic duality,
for any electric configuration (\ref{ep}), 
one can construct directly the dual magnetic solution.
This has (with $\ell>1$):
\begin{eqnarray}
A=\Phi(r,\theta)d \varphi,~~{\rm with}~~~
\Phi_\ell(r,\theta)=P_\ell(r)U_\ell( \theta),
\end{eqnarray}
where
 \begin{eqnarray}
P_\ell(r)=r^2 \frac{dR_\ell(r)}{dr},~~U_\ell( \theta)=\sin \theta \frac{d}{d\theta} \mathcal{P}_\ell(\cos \theta).
\end{eqnarray}


 \begin{small}
 
 \end{small}

\end{document}